\documentclass[12pt,twoside]{article}

\usepackage{pslatex}	
\usepackage{amsfonts}
\usepackage{amsmath}
\usepackage{amsthm}
\usepackage{amscd}
\usepackage{cite}
\usepackage{epsf}
\usepackage{a4}

\def\beq{\begin{equation}}
\def\eeq{\end{equation}}
\def\bea{\begin{eqnarray}}
\def\eea{\end{eqnarray}}

\def\beann{\begin{eqnarray*}}
\def\eeann{\end{eqnarray*}}

\let\a=\alpha

  \let\la=\lambda \let\m=\mu
   \let\r=\rho 
\let\om=\omega 
 \let\Ph=\phi  
  
\let\La=\Lambda  

\let\qd=\quad  

\def\epp{\, .}
\def\epc{\, ,}

\def\tst#1{{\textstyle #1}}
\def\dst#1{{\displaystyle #1}}
\def\sst#1{{\scriptstyle #1}}

\theoremstyle{plain}

\newtheorem*{corollary*}{Corollary}

\theoremstyle{definition}

\def\2{\frac{1}{2}} \def\4{\frac{1}{4}}

\def\6{\partial}

\def\+{\dagger}

\def\<{\langle} \def\>{\rangle}

 \let\ab=\downarrow

\def\i{{\rm i}}

\renewcommand{\appendix}{%
   \renewcommand{\section}{
	\secdef\Appendix\sAppendix}%
   \setcounter{section}{0}%
   \renewcommand{\thesection}{\Alph{section}}%
   \renewcommand{\theequation}{\thesection.\arabic{equation}}%
}

\newcommand{\Appendix}[2][?]{%
     \refstepcounter{section}%
     \setcounter{equation}{0}%
     \addcontentsline{toc}{appendix}%
          {\protect\numberline{\appendixname~\thesection} #1}%
     \vspace{\baselineskip}%
     {\noindent\large\bfseries\appendixname\ \thesection: #2\par}%
     \sectionmark{#1}\vspace{\baselineskip}}

\newcommand{\sAppendix}[1]{%
     {\noindent\large\bfseries\appendixname\: #1\par}%
     \sectionmark{#1}\vspace{\baselineskip}}

\begin{document}

\thispagestyle{empty}

\begin{center}

{\Large {\bf A note on the Bethe ansatz solution of the
supersymmetric ${\mathbf t}$-${\mathbf J}$ model\\}}

\vspace{7mm}

{\large Frank G\"{o}hmann\footnote[2]{e-mail:
goehmann@physik.uni-wuppertal.de} and Alexander Seel\footnote{e-mail:
seel@physik.uni-wuppertal.de}\\

\vspace{5mm}

Fachbereich Physik, Bergische Universit\"at Wuppertal,\\
42097 Wuppertal, Germany\\}

\vspace{20mm}

{\large {\bf Abstract}}

\end{center}

\begin{list}{}{\addtolength{\rightmargin}{10mm}
               \addtolength{\topsep}{-5mm}}
\item
The three different sets of Bethe ansatz equations describing the
Bethe ansatz solution of the supersymmetric $t$-$J$ model are
known to be equivalent. Here we give a new, simplified proof of
this fact which relies on the properties of certain polynomials.
We also show that the corresponding transfer matrix eigenvalues agree.
\\[2ex]
{\it PACS: 05.50.+q, 71.10.Fd, 71.10.Pm}
\end{list}

\clearpage

\section{Introduction}
The supersymmetric $t$-$J$ model, defined by the Hamiltonian
\begin{equation}
     H P_0 = P_0 \Bigl\{ - \sum_{j=1}^L
        (c_{j, a}^\+ c_{j+1, a} + c_{j+1, a}^\+ c_{j, a})
	+ 2 \sum_{j=1}^L \left( S_j^\a S_{j+1}^\a
	- \frac{n_j n_{j+1}}{4} + n_j \right) \Bigr\} P_0 \epc
\end{equation}
is the familiar $t$-$J$ model of solid state physics (for details see
e.g.\ \cite{Auerbach94}) at a certain ratio of hopping strength and
coupling, $t/J = \2$. For this ratio the Hamiltonian is invariant under
the action of the Lie super algebra gl(1$|$2) \cite{Wiegmann88,%
Foerster89}. Its one-dimensional version is solvable by Bethe
ansatz \cite{Schlottmann87} and is related to the lattice gas models
of Lai and Sutherland \cite{Lai74,Sutherland75}. An algebraic Bethe
ansatz solution was obtained by Essler and Korepin in \cite{EsKo92}
(see also \cite{FoKa93}).

A peculiarity of the model is the non-uniqueness of the form of the
Bethe ansatz solution. In fact, three different sets of Bethe ansatz
equations and three different expressions for the eigenvalue of the
associated transfer matrix were obtained in \cite{EsKo92}. The authors
of \cite{EsKo92} then used a technique \cite{Woynarovich83b} based on
the residue theorem in order to show the mutual equivalence of the three
sets of Bethe ansatz equations.

In a recent article \cite{GoSe03app} we reconsidered the Bethe ansatz
solution of the supersymmetric $t$-$J$ model. Our intention was to
give an example of how to apply the Bethe ansatz solution
\cite{GoSe03app} of the so-called gl(1$|$2) generalized model
\cite{Goehmann02}. We showed how the different gradings leading to the
different forms of the Bethe ansatz equations relate to different
choices of reference states for the Bethe ansatz. We found that if we
count the different ways to construct the eigenvectors, then there are
six different algebraic Bethe ansatz solutions of the supersymmetric
$t$-$J$ model which are pairwise related by spin-flip transformations.
From our considerations it also followed that the different expressions
for the Bethe ansatz eigenvalues must agree. Here we give a simple
direct proof of this fact, which is based on the properties of
polynomials rather than on integration in the complex plane. Our
technique also allows us to give a new proof of the equivalence of the
Bethe ansatz equations. It seems to be widely applicable and applies,
for instance, also to showing the relation between the eigenstates in
the repulsive and attractive cases of the Hubbard model
\cite{Woynarovich83b}.

\section{Equivalence of the three different forms of the Bethe ansatz
equations}
We show that the three different forms of the Bethe ansatz equations
for the supersymmetric $t$-$J$ model associated with the gradings
$(-,+,-)$, $(-,-,+)$ and $(+,-,-)$ are mutually equivalent.

Let us start with the case $(-,+,-)$. Then the Bethe ansatz equations
for twisted boundary conditions (twist angles $\Ph_s$ and $\Ph_c$) are
\begin{subequations}
\label{baempm}
\begin{align} \label{balam}
     & \biggl( \frac{\la_\ell - \tst{\frac{\i}{2}}}
                   {\la_\ell + \tst{\frac{\i}{2}}} \biggr)^L
        = e^{- \i \Ph_+} \prod_{j=1}^M
	   \frac{\la_\ell - \om_j - \tst{\frac{\i}{2}}}
	        {\la_\ell - \om_j + \tst{\frac{\i}{2}}} \epc \qd
		\ell = 1, \dots, N \epc \\[1ex] \label{baom}
     & \prod_{\ell = 1}^N \frac{\om_j - \la_\ell - \tst{\frac{\i}{2}}}
                               {\om_j - \la_\ell + \tst{\frac{\i}{2}}}
        = e^{- \i \Ph_c} \epc \qd j = 1, \dots, M \epc
\end{align}
\end{subequations}
where $\Ph_+ = \Ph_c + \Ph_s$. We shall further assume that $N \ge M$.
Equations (\ref{baempm}) turn into equations (4.3) of Essler and
Korepin for $\Ph_c = \Ph_s = 0$ if we identify $\la_k^{(1)} = \om_k$,
$\tilde \la_\ell = \la_\ell$, $N_\ab = M$, and $N_h + N_\ab = N$.

We define the polynomial
\begin{equation} \label{pf1}
     p(z) = \prod_{n=1}^N (z - \la_n - \tst{\frac{\i}{2}})
            - e^{- \i \Ph_c}
	    \prod_{n=1}^N (z - \la_n + \tst{\frac{\i}{2}}) \epp
\end{equation}
Let us fix a solution $\{ \{\la_\ell\}_{\ell = 1}^N, \{\om_j\}_{j=1}^M
\}$ of (\ref{baempm}). Then (\ref{baom}) implies that $p(\om_j) = 0$
for $j = 1, \dots, M$. $p$ is a polynomial of degree $N \ge M$. Besides
the zeros $\om_j$ it has $N - M$ additional zeros $\m_k$, $k = 1, \dots,
N - M$. Thus, $p(z)$ can be represented as
\begin{equation} \label{pf2}
     p(z) =(1 - e^{- \i \Ph_c}) \prod_{j=1}^M (z - \om_j)
	                        \prod_{k=1}^{N-M} (z - \m_k) \epp
\end{equation}
Comparing (\ref{pf1}) and (\ref{pf2}) we obtain
\begin{align}
     p(\la_\ell - \tst{\frac{\i}{2}}) & =
        \prod_{n=1}^N (\la_\ell - \la_n - \i) =
        (1 - e^{- \i \Ph_c})
	\prod_{j=1}^M (\la_\ell - \om_j -\tst{\frac{\i}{2}})
	\prod_{k=1}^{N-M} (\la_\ell - \m_k - \tst{\frac{\i}{2}}) \epc
	\notag \\
     p(\la_\ell + \tst{\frac{\i}{2}}) & =
        - e^{- \i \Ph_c}
        \prod_{n=1}^N (\la_\ell - \la_n + \i) =
        (1 - e^{- \i \Ph_c})
	\prod_{j=1}^M (\la_\ell - \om_j +\tst{\frac{\i}{2}})
	\prod_{k=1}^{N-M} (\la_\ell - \m_k + \tst{\frac{\i}{2}}) \epp
	\notag
\end{align}
Dividing the latter two equations we arrive at the identity
\begin{equation}
        e^{\i \Ph_c}
        \prod_{\substack{n = 1 \\ n \ne \ell}}^N
	   \frac{\la_\ell - \la_n - \i}{\la_\ell - \la_n + \i}
	\prod_{k=1}^{N-M} \frac{\la_\ell - \m_k + \tst{\frac{\i}{2}}}
	                       {\la_\ell - \m_k - \tst{\frac{\i}{2}}} =
	\prod_{j=1}^M \frac{\la_\ell - \om_j -\tst{\frac{\i}{2}}}
	                   {\la_\ell - \om_j +\tst{\frac{\i}{2}}}
\end{equation}
which, when inserted into the right hand side of (\ref{balam}), implies
\begin{subequations}
\label{baemmp}
\begin{align}
     & \biggl( \frac{\la_\ell - \tst{\frac{\i}{2}}}
                    {\la_\ell + \tst{\frac{\i}{2}}} \biggr)^L =
        e^{- \i \Ph_s}
        \prod_{\substack{n = 1 \\ n \ne \ell}}^N
	   \frac{\la_\ell - \la_n - \i}{\la_\ell - \la_n + \i}
	\prod_{k=1}^{N-M} \frac{\la_\ell - \m_k + \tst{\frac{\i}{2}}}
	                       {\la_\ell - \m_k - \tst{\frac{\i}{2}}}
			       \epc \qd \ell = 1, \dots, N \epp
\intertext{Moreover, since the $\m_k$ are zeros of $p(z)$ we conclude
with (\ref{pf1}) that}
     & \prod_{\ell = 1}^N \frac{\m_k - \la_\ell - \tst{\frac{\i}{2}}}
                               {\m_k - \la_\ell + \tst{\frac{\i}{2}}} =
       e^{- \i \Ph_c} \epc \qd k = 1, \dots, N - M \epp
\end{align}
\end{subequations}
Equations (\ref{baemmp}) are the Bethe ansatz equations for $(-,-,+)$
grading. Upon identifying $\tilde \la_\ell = \la_\ell$, $\tilde
\la_k^{(1)} = \m_k$ and setting $\Ph_s = \Ph_c = 0$ they turn into
equation (3.73) of \cite{EsKo92}. The polynomial $p(z)$ is fixed for
fixed $\{\la_\ell\}_{\ell = 1}^N$. Thus, every solution of
(\ref{baempm}) gives a solution of (\ref{baemmp}) and vice versa.

We may now apply the same polynomial trick to equation (\ref{balam}).
Let us define
\begin{equation} \label{qf1}
     q(z) = (z - \tst{\frac{\i}{2}})^L
            \prod_{n=1}^M (z - \om_n + \tst{\frac{\i}{2}}) -
	    e^{- \i \Ph_+}
	    (z + \tst{\frac{\i}{2}})^L
            \prod_{n=1}^M (z - \om_n - \tst{\frac{\i}{2}}) \epp
\end{equation}
The polynomial $q$ has $L + M$ zeros, the $\la_\ell$, $\ell = 1, \dots,
N$, are among them. Let us denote the remaining zeros by $\r_k$,
$k = 1, \dots, L - N + M$ (we assume that $N \le L$). Then
\begin{equation} \label{qf2}
     q(z) = (1 - e^{- \i \Ph_+}) \prod_{\ell = 1}^N (z - \la_\ell)
            \prod_{k=1}^{L - N +M} (z - \r_k) \epp
\end{equation}
Evaluating $q(z)$ at $ z = \om_j - \tst{\frac{\i}{2}}$ and at
$z = \om_j + \tst{\frac{\i}{2}}$ by using both forms, (\ref{qf1}) and
(\ref{qf2}), of the polynomial and dividing the resulting equations by
each other we obtain the identity
\begin{equation}
     e^{- \i \Ph_+} \prod_{\substack{n = 1 \\ n \ne j}}^M
        \frac{\om_j - \om_n - \i}{\om_j - \om_n + \i}
        \prod_{k=1}^{L-N+M}
        \frac{\om_j - \r_k + \tst{\frac{\i}{2}}}
             {\om_j - \r_k - \tst{\frac{\i}{2}}} =
        \prod_{\ell = 1}^{N}
        \frac{\om_j - \la_\ell - \tst{\frac{\i}{2}}}
             {\om_j - \la_\ell + \tst{\frac{\i}{2}}}
\end{equation}
which holds for $j = 1, \dots, M$. Inserting this identity into
(\ref{baom}) it follows that
\begin{subequations}
\label{baepmm}
\begin{equation}
     \prod_{k=1}^{L-N+M} \frac{\om_j - \r_k - \tst{\frac{\i}{2}}}
                              {\om_j - \r_k + \tst{\frac{\i}{2}}} =
     e^{- \i \Ph_s} \prod_{\substack{n = 1 \\ n \ne j}}^M
        \frac{\om_j - \om_n - \i}{\om_j - \om_n + \i} \epc
	\qd j = 1, \dots, M \epp
\end{equation}
Since the $\r_k$ satisfy $q(\r_k) = 0$, we also have
\begin{equation}
     \biggl( \frac{\r_k - \tst{\frac{\i}{2}}}{\r_k + \tst{\frac{\i}{2}}}
        \biggr)^L = e^{- \i \Ph_+} \prod_{j=1}^M
	              \frac{\r_k - \om_j - \tst{\frac{\i}{2}}}
	                   {\r_k - \om_j + \tst{\frac{\i}{2}}} \epc
			   \qd k = 1, \dots, L - N + M \epp
\end{equation}
\end{subequations}
Equations (\ref{baepmm}) are the Bethe ansatz equations for the grading
$(+,-,-)$. We recover equation (3.49) of \cite{EsKo92} setting
$\tilde \la_k = \r_k$, $\la_j^{(1)} = \om_j$, $\Ph_s = \Ph_c = 0$,
$L - N + M = N_e$, $M = N_\ab$.

To sum up, we have shown the mutual equivalence of the three sets of
Bethe ansatz equations (\ref{baempm}), (\ref{baemmp}) and (\ref{baepmm})
by exploiting the properties of the polynomials $p(z)$ and $q(z)$
defined in equations (\ref{pf1}) and (\ref{qf1}).

\section{Equivalence of the transfer matrix eigenvalues}
The polynomials $p(z)$ and $q(z)$ also allow us to show that the
eigenvalues corresponding to the three forms of the Bethe ansatz
solutions agree.

The transfer matrix eigenvalue for the grading $(-,+,-)$ is
\begin{multline} \label{evmpm}
     \La_{(-+-)} (\la) = - \biggl( \frac{\la - \i}{\la + \i} \biggr)^L
        \prod_{\ell = 1}^N \frac{\la - \la_\ell + \tst{\frac{\i}{2}}}
                                {\la - \la_\ell - \tst{\frac{\i}{2}}}
        \\ + \biggl( \frac{\la}{\la + \i} \biggr)^L
          \biggl( \prod_{\ell = 1}^N
	     \frac{\la - \la_\ell + \tst{\frac{\i}{2}}}
	          {\la - \la_\ell - \tst{\frac{\i}{2}}} - 1 \biggr)
        \prod_{j=1}^M \frac{\la - \om_j - \i}{\la - \om_j} \epc
\end{multline}
where the Bethe ansatz roots $\la_\ell$ and $\om_j$ are solutions of
(\ref{baempm}) with $\Ph_s = \Ph_c = 0$. $\La_{(-+-)}$ agrees with
the expression (4.4) of \cite{EsKo92} if we set $\la_j^{(1)} = \om_j$,
$\tilde \la_\ell = \la_\ell$, $N_\ab = M$ and $N_h + N_\ab = N$.

It follows from equations (\ref{pf1}) and (\ref{pf2}) that
\begin{equation}
     \frac{p(\la - \i)}{p(\la)} =
        \prod_{j=1}^M \frac{\la - \om_j - \i}{\la - \om_j}
        \prod_{k=1}^{N-M} \frac{\la - \m_k - \i}{\la - \m_k}
	= \frac{e^{- \i \Ph_c} - \dst{\prod_{\ell = 1}^N}
	             \frac{\la - \la_\ell - \sst{\frac{3 \i}{2}}}
		          {\la - \la_\ell - \sst{\frac{\i}{2}}}}
	         {e^{- \i \Ph_c} \dst{\prod_{\ell = 1}^N}
	       \frac{\la - \la_\ell + \sst{\frac{\i}{2}}}
	            {\la - \la_\ell - \sst{\frac{\i}{2}}} - 1} \epp
\end{equation}
Setting $\Ph_c = 0$ we obtain the identity
\begin{equation}
     \biggl( \prod_{\ell = 1}^N
	       \frac{\la - \la_\ell + \tst{\frac{\i}{2}}}
	            {\la - \la_\ell - \tst{\frac{\i}{2}}} - 1 \biggr)
             \prod_{j=1}^M \frac{\la - \om_j - \i}{\la - \om_j}
	= \biggl(1 - \prod_{\ell = 1}^N
	             \frac{\la_\ell - \la + \tst{\frac{3 \i}{2}}}
		          {\la_\ell - \la + \tst{\frac{\i}{2}}} \biggr)
        \prod_{k=1}^{N-M} \frac{\la - \m_k}{\la - \m_k - \i} \epp
\end{equation}
This identity allows us to replace the $\om_j$ in equation (\ref{evmpm})
in favour of the Bethe ansatz roots $\m_k$ connected with $(-,-,+)$
grading. We obtain
\begin{multline} \label{evmmp}
     \La_{(-+-)} (\la) = - \biggl( \frac{\la - \i}{\la + \i} \biggr)^L
        \prod_{\ell = 1}^N \frac{\la - \la_\ell + \tst{\frac{\i}{2}}}
                                {\la - \la_\ell - \tst{\frac{\i}{2}}}
        \\ + \biggl( \frac{\la}{\la + \i} \biggr)^L
	\biggl(1 - \prod_{\ell = 1}^N
	           \frac{\la_\ell - \la + \tst{\frac{3 \i}{2}}}
		        {\la_\ell - \la + \tst{\frac{\i}{2}}} \biggr)
        \prod_{k=1}^{N-M} \frac{\la - \m_k}{\la - \m_k - \i} \epp
\end{multline}
But the latter expression is precisely the expression for the
eigenvalue $\La_{(--+)} (\la)$ (which has to be compared with (3.74)
of \cite{EsKo92}). We may formulate this result in the following way:
\begin{equation}
     \La_{(-+-)} \bigl( \la \big| \{\la_\ell\}_{\ell = 1}^N,
                                  \{\om_j\}_{j=1}^M \bigr) =
     \La_{(--+)} \bigl( \la \big| \{\la_\ell\}_{\ell = 1}^N,
                                  \{\m_k\}_{k=1}^{N-M} \bigr) \epc
\end{equation}
where we exposed explicitly the dependence of the transfer matrix
eigenvalues on the Bethe ansatz roots. The two sets of Bethe ansatz
roots $\{\om_j\}_{j=1}^M$ and $\{\m_k\}_{k=1}^{N-M}$ are connected by
the polynomial $p(z)$, equation (\ref{pf1}).

In a similar way $\La_{(-+-)}$ is related to $\La_{(+--)}$ through the
polynomial $q(z)$. The proof should now be obvious and is left as an
exercise to the reader.


\providecommand{\bysame}{\leavevmode\hbox to3em{\hrulefill}\thinspace}
\providecommand{\MR}{\relax\ifhmode\unskip\space\fi MR }
\providecommand{\MRhref}[2]{%
  \href{http://www.ams.org/mathscinet-getitem?mr=#1}{#2}
}
\providecommand{\href}[2]{#2}

\end{document}